# Spike-timing-dependent-plasticity learning in a planar magnetic domain wall artificial synapse

J. O. Castro,[1] B. Buyatti,[1] D. Mercado,[2, 3] A. Di Donato,[2, 3] M. Quintero,[2, 4] and M. Tortarolo[2, 5, *]

[1] *Universidad de Buenos Aires (UBA), Facultad de Ciencias Exactas y Naturales (FCEN)*
[2] *Instituto de Nanociencia y Nanotecnología (CNEA - CONICET), Nodo Buenos Aires,*
*Av. Gral. Paz 1499, B1650 Villa Maipú, Provincia de Buenos Aires, Argentina*
[3] *Departamento de Micro y Nanotecnología, GDTyPE-GAIDI-CNEA,*
*Av., General Paz 1499, (1650) San Martín, Buenos Aires, Argentina*
[4] *Departamento de Física de la Materia Condensada (GAIDI-CNEA),*
*Av. Gral. Paz 1499, B1650KNA San Martín, Prov. Buenos Aires, Argentina*
[5] *Laboratorio Argentino de Haces de Neutrones (CNEA),*
*Centro Atómico Constituyentes, Av. Gral. Paz 1499,*
*B1650 Villa Maipú, Provincia de Buenos Aires, Argentina*

Future neuromorphic architectures will require millions of artificial synapses, making it mandatory to understand the physical mechanisms behind their plasticity functionalities. In this work, we propose a simplified spin memristor, where the resistance can be controlled by magnetic field pulses, based on a Co/Pt multilayer with perpendicular magnetic anisotropy as a synapsis emulator. We demonstrate plasticity and spike time dependence plasticity (STDP) in this device and explored the underlying magnetic mechanisms using Kerr microscopy imaging and Hall magneto-transport measurements. A well-defined threshold for magnetization reversal and the continuous resistance states associated with the micromagnetic configuration are the basic properties allowing plasticity and STDP learning mechanisms in this device.

## I. INTRODUCTION

The ability to learn in a biologically inspired artificial synapse relies on its capacity to reconfigure in response to a stimulus, weakening or strengthening its synaptic weight. This property is called plasticity, and it is the basis of one of the most representative learning rules, the spike-timing-dependent plasticity STPD [1–4]. This rule enables unsupervised learning and relies on the relative timing and causality in which the action potentials from pre- and post-synaptic neurons are emitted [5]. If the temporal difference between pre-and post-synaptic pulse $\Delta t$ is positive, the synaptic weight strengthens, and if it is negative, the synaptic weight weakens.

During the last decade, the emulation of artificial synapses with memristor devices has been explored [6–8]. The resistance of these devices depends on the history of electrical signals that were previously applied, having memory properties [9]. The memristor resistance is inversely proportional to the synaptic weight, the resistance decreases (synaptic potentiation) if $\Delta t > 0$ and it increases if $\Delta t < 0$ (synaptic depression). Several device implementations have been proposed such as phase change[10], oxide-based resistive switching [11–13], and spin memristors [14–20]. In a spin memristor, the magnetoresistance is related to the domain wall (DW) dynamics or, in general, to the magnetization reversal mechanisms at play. The main aspect of their memristive functionality is the magnetization control with either magnetic field [21], current [15] or electric field [22], being analog and non-volatile. Hwang et al [21] proposed a magnetic field-driven DW device in a multiple Hall cross (MHCs) geometry, showing its memristive properties and focusing on the demonstration of the multilevel programming capacities of this relatively simple fabrication device, compared to the magnetic tunnel junctions (MTJ) devices usually characterized for these applications [14, 17]. It was implemented using field-driven DW motion, but the results directly apply to current-driven DW motion [20].

Inspired by their simplified device, in this work, we demonstrate plasticity and STPD implementation in a planar single-Hall-cross device on the second timescale, nevertheless, results are valid at the ns scale due to the characteristic DW velocity [23–26]. Our device was fabricated from a Co/Pt thin film multilayer, studied for decades due to its large perpendicular magnetic anisotropy (PMA)[27–30]. Also, this material system is well known in the field of ultrafast magnetism that might contribute to the development of future spintronic and data storage applications, with an ultrafast magnetization quenching on timescales $\Delta \ll 1$ ps [31, 32]. This single hall-cross design can contribute to the integration of spin memristors in neuromorphic applications, especially due to its simple fabrication process, being a fast bench test for neuromorphic properties of novel neuromorphic architectures.

## II. SAMPLE AND DEVICE CHARACTERIZATION

Our device is based on a sputter-deposited Si/[Pt (20 A)/ Co (4 A)]$_{\times 6}$/ Pt (30 A)) multilayer structure. The [Co/Pt] film was patterned into a simple Hall cross of

---

* Corresponding author: marinatortarolo@cnea.gob.ar

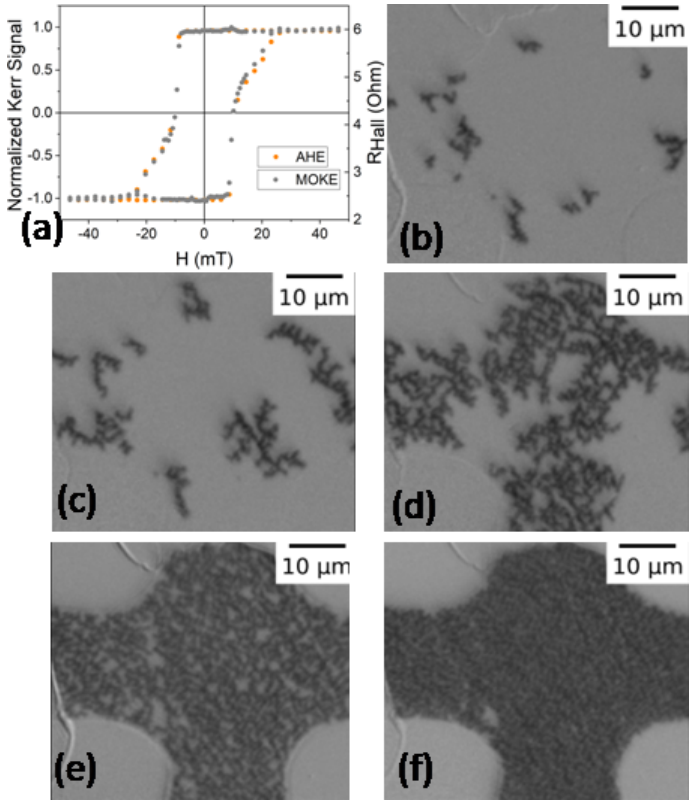

FIG. 1. (color online) (a) Hysteresis loop for MOKE (grey dots) and anomalous Hall effect (AHE) magnetoresistance (orange dots) measurements. Kerr microscopy images of the magnetic domain nucleation and propagation for (b) -8 mT, (c) -9 mT, (d) -9.5 mT (e) -10 mT, and -13 mT(f).

30 $\mu m^2$ area using photo-lithography followed by $Ar^+$ ion milling. 4 Au pads were defined by lift-off and wire bonded to the sample socket to perform the magneto transport measurements. Device resistance was measured using a 4-point probe, with a 4 mA probing current, and the measured voltages were recorded to calculate R = $\Delta V/I$.

The sample was magnetically characterized by static magneto-optic Kerr effect (MOKE) and the Hall magnetoresistance ($R_{Hall}$) measurements simultaneously (Fig.1 (a)), showing the characteristic PMA of this kind of [Co/Pt] multilayers with less than 1 nm Co thickness[27–30] with a coercive field $H_c \sim 10$ mT, with a high resistance state (HRS) of 6 $\Omega$ and a low resistance state (LRS) of 2.5 $\Omega$. Both MOKE and resistance were taken in remanence. The anomalous Hall magnetoresistance $R_{Hall}$ is related to the perpendicular magnetization $M_z$ of the sample by the second term of:

$$R_{\text{Hall}} = R_o H_z + R_A M_z(T, H_z). \qquad (1)$$

Where $R_o$ is the ordinary Hall coefficient, $M_z$ is the magnetization parallel to the z-axis, $H_z$ is the external magnetic field, and $R_A$ is the anomalous Hall coefficient.

As our measurements were taken in remanence ($H_z$), we are measuring the anomalous Hall effect (AHE), represented by the second term of eq. (1). The overlap of MOKE and Hall resistivity signal shows that the resistance of the device is proportional to the magnetization of the sample (Fig.1 (a)) and validates the use of the $R_{Hall}$ as the device read-out signal.

We used Kerr microscopy imaging to correlate the micromagnetic reversal with $R_{Hall}$ ( see Fig.1 (a)). The sample magnetization was saturated at 45 mT before each image was acquired, subsequently applying 0.2 s magnetic field pulses of (b) -8 mT, (c) -9 mT, (d) -9.5 mT, (e) -10 mT and (f) -13 mT. The dominant magnetization reversal mechanism for lower fields up to $\sim$ 9 mT is the nucleation of reversed magnetic domains. Above 9 mT, the dominant reversal mechanism is the domain propagation with dendritic shape up to the complete magnetization reversal at $\sim$ 30 mT [33–35]. Kerr images reveal that the magnetic domain nucleation and propagation produce several intermediate magnetic and resistive states that can be controlled by magnetic field pulses. These states are the microscopic mechanism behind memristive and plasticity properties of the device. Notice that the relatively high density of magnetization reversal domains enables a further reduction in the lateral sizes of the Hall cross structure, even up to the nanoscale [23].

To characterize plasticity properties, the device was subject to a train of pulses of 0.2 s duration of increasing/decreasing amplitude (Fig. 3 (a)) and of constant amplitude (Fig. 3 (b)). This pulse duration was chosen according to the rise time of the magnetic field coil. The $R_{Hall}$ was measured immediately after each field pulse to determine the resistive state.

The pre and post-neuron spikes (action potentials) were modeled by two identical waveforms presented in Fig.4 (a) and (b), consisting of rectangular voltage shapes followed by smooth slopes of opposite polarity, but several other pre- and post-spike shapes are proposed in the literature [6, 36]. The overlapping of both pre- and post-neuron spikes within a $\Delta$t delay produces a waveform ($V_{pre}$-$V_{post}$) displayed in Fig.4, considering a positive delay (c) and a negative delay (e). Each waveform was used to source the current to the magnetic field coil thus modulating the magnetic state with an applied magnetic field, and the resistance of the device was probed after each magnetic field pulse (t= 0.2 s) for each delay Fig.4 (d and f). A 0-40 s delay sequence was measured, considering positive and negative $\Delta$t. Notice that each spike individually does not exceed the $H_c$, hence the resistance value is modulated by the amplitude and polarity of the $V_{pre}$-$V_{post}$ waveform. The synaptic weight $\Delta$w for each delay is defined as the ratio of the initial resistance to the final resistance after the spiking (Fig.4 d and f).

## III. PLASTICITY AND STDP RESULTS AND ANALYSIS

The hysteresis loop in Fig. 1 (a) shows a clear threshold around the $+H_c$ ($-H_c$) beyond which magnetization switches, and increases (decreases) up to $+M_{sat}$ ($-M_{sat}$) (HRS and LRS, respectively). Between these two states, the device shows a continuous resistance variation under H, originating from the micromagnetic domain nucleation and propagation shown by Kerr imaging (see Fig.1 (b,c,d,e,f)). These well-determined $H_c$ thresholds allow us to implement the STDP learning rule [37] in these magnetic field-driven devices, and the continuous resistance states are a requirement for the learning processes. Figure 2 shows the micromagnetic evolution of these continuous states under 0.2 s pulses of -9 mT (upper panel) and +9 mT (lower panel), similar to the ones applied in Figure 3 (b). The magnetization reversal proceeds fast up to around pulse 25, to start saturating to the maximal magnetization value corresponding to this applied field. The plateau in the $R_{Hall}$ signal around pulse 25 corresponds to this magnetic saturation, though, as shown in Fig. 5a (See Appndix A) magnetization is not completely reversed within the full span of the experiment. This could be attributed to pinning effects due to local defects on the magnetic layer. The micromagnetic image (Fig. 2, upper panel) at pulse 50 still shows non-reversed clear areas, possibly defects, that can be the cause of the distancing of the mean Kerr intensity measured (proportional to magnetization) from the predicted relaxation behavior of eq. A1 (See Appendix A) for longer cumulated pulse durations. Local defects can modify the magnetic anisotropy, generating hard magnetization centers that require either higher Zeeman energy or longer relaxation times to evolve into a uniform magnetic state [33]. Each time a magnetic field pulse, or the temporal overlap of two pulses separated by $\Delta t$, overcomes the threshold ($H_c$) the magnetization reversal leads to a resistive state that reconfigures for each magnetic field pulse, giving rise to the plasticity and the STDP learning rule discussed on the following paragraphs.

Synaptic plasticity refers to the ability of synapsis to adjust its synaptic weight in response to input stimuli. This feature was investigated by applying two types of spike stimulus: one with a ramp-like amplitude and one with a constant amplitude. Figure 3 (a) shows the device resistance for several depression (resistance increase)/potentiation (resistance decrease) cycles (upper panel) applying spike trains of fixed time width and increasing (decreasing) magnetic field (ramp-like amplitude). During synaptic depression, the spikes corresponding to lower magnetic fields (0- 9mT) do not induce any resistance changes up to a threshold that can be identified as the $H_c$. After surpassing this threshold, the resistance gradually increases mimicking the shape of the magnetic hysteresis loop. The device presents several intermediate resistive states between $H_c$ and $-H_c$ that can be associated with the micromagnetic structure evolution with H. In Figure 3 (b), plasticity is investigated through several depression/potentiation cycles of identical amplitude spikes. During both cycles, the resistance gradually varies showing several intermediate states between the HRS and LRS. In all the cycles the resistance value does not vary constantly with the spike number, but it changes notably faster for the first spikes in both depression and potentiation cycles. In both cases, ramp-like and constant amplitude, we observe a persistent strengthening/weakening of the synapses after stimuli, characteristic of long-term synaptic plasticity. Also, as observed in several potentiation and depression cycles in Fig.3, the device's intermediate resistive states and the HRS and LRS are reproducible. Overall, plasticity is directly related to the micromagnetic structure, as can be related from Fig. 2. In this sense, plasticity is enhanced above the coercive field in both pulse shapes studied (ramp-like and constant amplitude).

Based on these plasticity results on magnetic field modulation, we performed STDP experiments relying on the pre- and post-spike temporal overlap. The resistance in the memristor is modulated by the waveform resulting from the difference of the pre and post-spike (Fig. 4 (a and b), which depends on their relative timing. To emulate STDP with $\Delta t > 0$ ($\Delta t < 0$) the device is first brought to LRS by negatively saturating its magnetization. Then the pre-and post-spike difference waveforms are applied as a magnetic field. As the resulting waveforms Fig. 4 (c and e) temporarily exceed the $H_c$, i.e., the threshold of the system, leading to synaptic potentiation ($\Delta G > 0$) or a synaptic depression ($\Delta G < 0$), depending on the sign of $\Delta t$. Figure 4 (g) shows the STDP curve represented as the normalized conductance (G=1/R) change as a function of the spike delay, which has implicit the dependence on the intensity and the time interval of the applied magnetic field. The plot derived from the amplitude variation of the resistance qualitatively displays the same behavior as the biological STDP curve shown in Bi *et al*[5], showing that only closely timed spikes produce a conductance change on the device, while long delays do not impact on the conductance. This STDP behavior is characteristic of associative learning processes.

## IV. CONCLUSIONS

In summary, we studied the synaptic characteristics of a multistate magnetic field-driven domain wall memristor. In this device, when the magnetic field pulses overcome the $H_c$ threshold the magnetization reversal propagation gives rise to a continuum of resistive states showing an analogic and reproducible resistance variation within a 5 $\Omega$ range, that can be reconfigured by subsequent magnetic field pulses. These resistance states were associated with the micromagnetic dynamics of the device, which is also the underlying mechanism leading to long-term plasticity. The plasticity properties of the device were studied through stimulation with two different waveforms: ramp-like and constant, showing in both cases its capacity to emulate long-term potentiation and depression. Also, when two sub-threshold magnetic pulses separated by a Δt interval temporarily overlap producing a supra-threshold pulse, the magnetization reversal dynamics enable the associative learning STDP rule, showing a biological-like behavior. This device, characterized by its extremely simplified fabrication process can serve as a fast test bench for neuromorphic computing architectures based on domain wall devices.

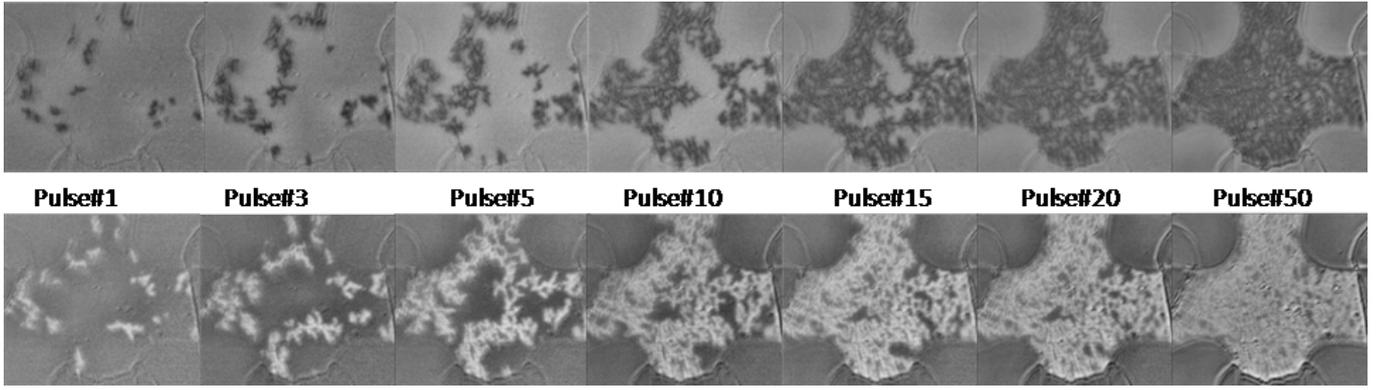

FIG. 2. Kerr microscopy images of the magnetic domain evolution for a series of 0.2 s magnetic field pulses of -9 mT (upper panel) and 9 (mT) (lower panel). The image scale is the same as in Fig.1 The sample was magnetically saturated with the opposite magnetization before each pulse series.

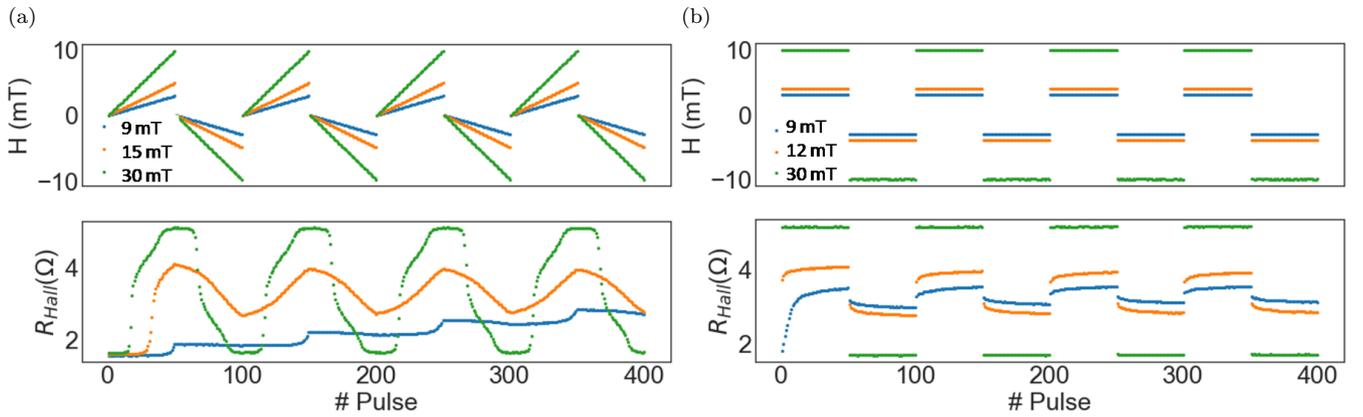

FIG. 3. Plasticity under different waveforms: (a) Top panel: Potentiation and depression cycles under ramp-like spike trains, Spike time width = 0.2 s, Depression ramps: 0 to 9mT, 0 to 15 mT, and 0 to 30 mT, potentiation ramps: 0 to -9 mT, 0 to -15 mT, 0 to -30 mT. Bottom panel: device resistance after each spike. (b) Top panel: Potentiation and depression cycles using identical 0.2 s spike trains for several amplitudes, depression: 9 mT, 12 mT, and 30 mT, potentiation: -9 mT, -12 mT, -30 mT. Bottom panel: device resistance after each spike.

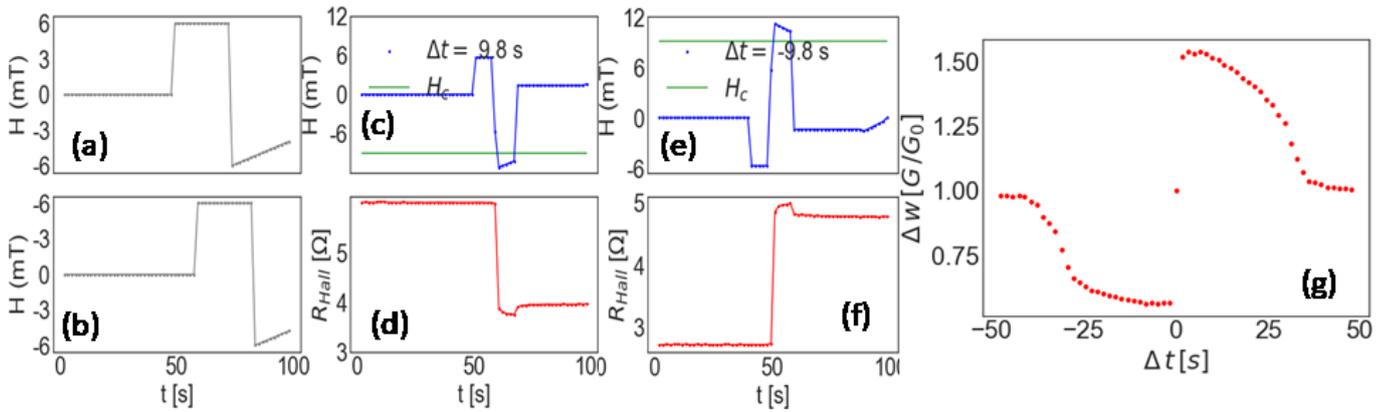

FIG. 4. (a) Waveforms used as pre- (b) and post-spike for the STDP protocol. Overlapping of the pre-and post-spike yielding to (c) potentiation and (e) depression, and their respective corresponding resistance (d) drop or (f) increase. (g) STDP learning curve.

## ACKNOWLEDGMENTS

We acknowledge P. Levy for the fruitful discussions and careful reading of this manuscript. The magnetic films were deposited and characterized at the Magnetron MOKE@LCP-MR facility supported by Sorbonne Université and the CNRS. L. A. Rosellini and M. Tracchia participated in the experimental setup. PIP-CONICET 11220210100516CO and PICT-2021-GRF-TII-00403 supported this work.

## Appendix A: Magnetization reversal-Plasticity fit

The magnetization reversal with the applied field pulses was obtained from the mean intensity of each Kerr image (Fig. 5a). Both curves corresponding to + 9 mT and 9 mT applied fields were fit with the usual logarithmic model from magnetic relaxation: $M(t) = M_o - S.\ln(1+t/t_o)$ (Eq. A1), where $M_o$ is the initial magnetization, S is the magnetic viscosity and $t_o$ is the characteristic time of the measurement[38]. In the case of the magnetic relaxation at - 9 mT, it was possible to fit only the 0 - 3 s interval by eq. A1. For longer cumulated pulse durations the measured magnetization deviates from the logarithmic behavior (simulated in violet dots using the parameters from the fit at the 0 - 3 s interval) possibly due to pinning effects in local defects. Using the same parameters obtained from eq. A1 fit of the magnetization, we fit the $R_{Hall}$ corresponding to the first plasticity cycle at 9 mT by the model $R_{Hall} = R_{HRS}(M_o - S.\ln(1+t/t_o)) + R_L RS[1-M_o - S.\ln(1+t/t_o)]$ (Eq. A2), describing quite satisfactorily the plasticity curve for the 0-3 s temporal interval (Fig. 5b). This fact highlights the origin of plasticity in micromagnetic states.

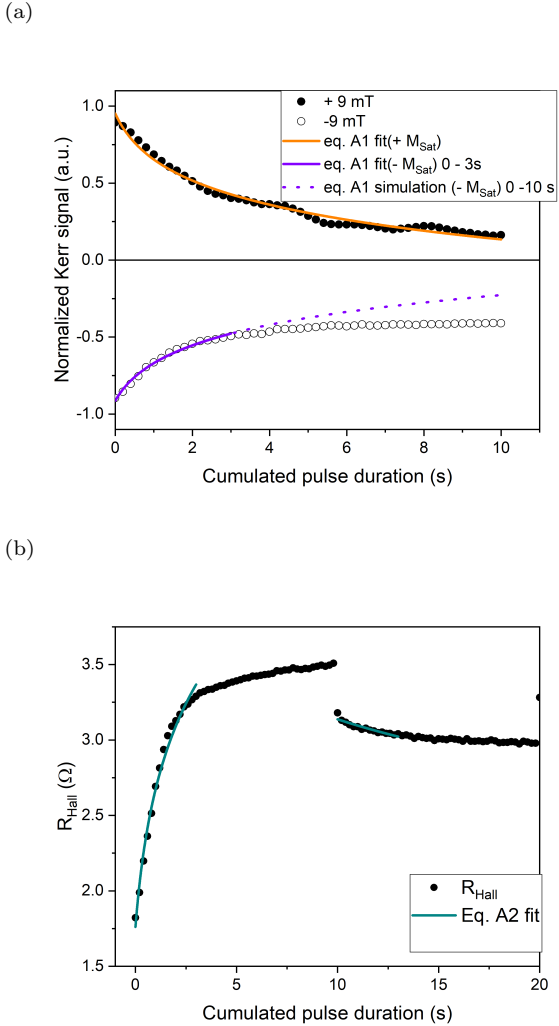

FIG. 5. (a) Mean magnetization extracted from the images in fig.2 as a function of the cumulated pulse duration for + 9 mT (filled black circles) and - 9 mT (empty black circles) fit by eq. A1. The dotted violet curve corresponds to a simulation for higher cumulated pulse durations. (b) $R_{Hall}$ signal corresponding to the first cycle of the 9 mT curve shown in Fig.3(b) fit by eq.A2.